\documentclass[11pt]{article}
\advance \topmargin by -\headheight
\advance \topmargin by -\headsep
\evensidemargin \oddsidemargin
\usepackage{hyperref}

\newcommand\nc{\newcommand} \nc\rnc{\renewcommand} 

\def\mboxes(#1){\xmboxes#1,xxx,}\def\endpiece{xxx}
\def\xmboxes#1,{\def\tmp{#1}%
\ifx\tmp\endpiece \else\expandafter\def\csname #1\endcsname{\hbox{#1}} 
\expandafter\xmboxes \fi}

\nc\nn{\newenvironment} \nc\nt{\newtheorem}
\nt{Corollary}{Corollary}[section]
\nt{Lemma}{Lemma}[section]
\nt{ProofAux}{Proof} 
\nn{Proof}{\begin{ProofAux}\em}{\ $\Box$\end{ProofAux}}

\nc\w{\wedge} \nc\st{\,:\:}

\nc\bs{\bigskip} \nc\ms{\medskip} \rnc\ss{\smallskip}
\rnc\ni{\noindent} \nc\ds{\displaystyle}

\nn{eatab}{\tt\begin{tabbing}mmm\=mmm\=mmm\=mmm\=mmm\=\kill}{\end{tabbing}}
   
\nc\false{\mbox{\em false\/}} \nc\true {\mbox{\em true\/}}
\nc\undef{\mbox{\em undef\/}}

\nc\bara{{\bar a}} \nc\bart{{\bar t}} \nc\barx{{\bar x}}

\nc\cA{{\cal A}} \nc\cS{{\cal S}}

\nc\al{\alpha} \nc\be{\beta} \nc\ga{\gamma} \nc\de{\delta}
\nc\Ups{\Upsilon}

\mboxes(Active,Bound,Dom,Fork,Free,Fun,Guard,Import,Input,Loc,Me,Mod,Moves,NUpdates,Prog,Reserve,Self,Terms,Type,Undecl,Updates,US,Val,Var,View)

\title{Evolving Algebras 1993:\ \ Lipari Guide\footnotemark[1]}

\author{\bf Yuri Gurevich\footnotemark[2]\footnotemark[3]}

\date{} 
\begin{document} \thispagestyle{plain} \maketitle

\footnotetext[1]{\copyright 1995 Oxford University Press.\quad
Published in
{\em Specification and Validation Methods\/}, 
Ed.\ E. B\"orger, Oxford University Press, 1995, 231--243.}

\footnotetext[2]{Partially supported by ONR grant N00014-91-J-1861 and
NSF grant CCR-92-04742.}

\footnotetext[3]{Soon after this article was published, evolving algebras were renamed abstract state machine because some clients were spooked by ``algebra''.}

\tableofcontents

\newpage
\section{Introduction}

Computation models and specification methods seem to be worlds apart.  The
evolving algebra project started as an attempt to bridge the gap by
improving on Turing's thesis \cite{G1,G2}.  We sought more versatile machines
which would be able to simulate arbitrary algorithms in a direct and
essentially coding-free way.  Here the term algorithm is taken in a broad
sense including programming languages, architectures, distributed and
real-time protocols, {\em etc.\/}.  The simulator is not supposed to
implement the algorithm on a lower abstraction level; the simulation should
be performed on the natural abstraction level of the algorithm.

The evolving algebra thesis asserts that evolving algebras are such
versatile machines.  The thesis suggests an approach to the notorious
correctness problem that arises in mathematical modeling of
non-mathematical reality: How can one establish that a model is faithful
to reality?  The approach is to construct an evolving algebra $\cal A$
that reflects the given computer system so closely that the correctness
can be established by observation and experimentation.  (There are tools
for running evolving algebras.)  $\cal A$ can then be refined or coarsened
and used for numerous purposes.  An instructive example is described in
\cite{B} by Egon B\"orger who championed this approach and termed $\cal A$
the {\em ground model\/} of the system.  The use of the successive
refinement method is facilitated by the ability of evolving algebras to
reflect arbitrary abstraction levels.  This has been convincingly
demonstrated by B\"orger and Rosenzweig in \cite{BR}; a simpler example is
found in \cite{GH}.

Evolving algebras have been used to specify languages ({\em e.g.\/} C,
Prolog and VHDL), to specify real and virtual architectures ({\em e.g.\/}
APE, PVM and Transputer), to validate standard language implementations
({\em e.g.\/} of Prolog, Occam), to validate distributed protocols (see
examples in Parts III and IV of this book), to prove complexity results
\cite{BG}, {\em etc.\/}.  See B\"orger's annotated bibliography on evolving
algebras in this book and the proceedings of the first evolving algebra
workshop in \cite{PS}.

Here we extend the definition of evolving algebras given in the tutorial
\cite{G2} (henceforth ``the tutorial'').  For the sake of brevity, the term
``evolving algebra'' is often shortened to ``ealgebra'' (pronounced
e-algebra) or ``EA''; the latter term is used mostly as an adjective.
Static algebras are discussed in \S2.  Sequential ealgebras are discussed
in \S3; first we define basic ealgebras and then we equip them with the
ability to import new elements.  Nondeterministic sequential ealgebras and
some other simple extensions of basic ealgebras are discussed in \S4,
parallel ealgebras are discussed in \S5, and distributed ealgebras are
discussed in \S6 which can be read immediately after \S3.  Admittedly this
guide is harder to read than the tutorial, and we intend to write a more
popular version of the guide.  

Now let us return to the EA thesis.  In the tutorial, we defined sequential
ealgebras and sketched a speculative philosophical ``proof'' of the
sequential version of the thesis.  The definition of sequential ealgebras
and the sequential EA thesis have survived several years of intensive
application and experimentation.  As a matter of fact, we (the EA
community) seem to have run out of challenges.

The situation with non-sequential computations is more complicated.  It
seems that, for every reasonably understood class of algorithms, there is a
natural extension of the basic EA model that ``captures'' that class.  That
form of the EA thesis also has survived several years of intensive
application and experimentation.  The philosophy and guiding principles of
the EA approach seem quite stable.  However, at the current stage of
computer science, there is yet no clear understanding of what parallel,
distributed or real-time algorithms are in general.  Thus, the definitions
of parallel and distributed ealgebras given below are necessarily
tentative.  They provide a foundation for existing EA applications and
reflect my anticipation of things to come.  (Many existing applications,
including those in this volume, were done before this guide have been
completed; the terminology there may reflect earlier versions of the
guide.)

We try to derive our definitions from first principles.  Unfortunately some
arbitrariness is inescapable and one has to balance the clarity and
simplicity versus programming convenience and efficient execution.  When
one thinks mostly about applications, as we do, there is a tendency to
prefer programming convenience and efficient execution.  This is a
dangerous trend which leads to an idiosyncratic programming language.  For
future reference we formulate the following principle:

\subparagraph{The Pragmatic Occam's Razor}
Logic simplicity comes first; it may be sacrificed only in those cases
where a slight logic complication is demonstrated to ease programming or
improve execution efficiency in a substantial way.\ss
  
The EA field is quickly expanding in depth and breadth.  I hope that this
guide lives up to its name and guides the developments in the near future.

\paragraph{Acknowledgment} 
Egon B\"orger and Dean Rosenzweig generously shared with me their ideas and
rich application experience.  Discussions with Andreas Blass were
indispensable in clarifying things.  Numerous working walks with Jim
Huggins through the woods of Ann Arbor were very helpful.  Raghu Mani
raised important implementation issues.  Numerous ealgebraists commented on
earlier drafts of the guide.  I am very thankful to all of them.  To an
extent, this chapter is a result of a collective effort, though I am
responsible for possible blunders.

A preliminary version of the guide has been tried out during the 1993
summer school on Specification and Validation Methods for Programming
Languages and Systems on the beautiful island of Lipari in Italy.  I use
this opportunity to thank the organizers, Egon B\"orger and Alfredo Ferro,
and all participants.

\section{Static Algebras and Updates}

\subsection{Static Algebras: Motivation}

In first-order logic, a structure is a nonempty set with operations and
relations (called the basic operations and relations of the structure).
That is how Tarski defined structures.  He could have defined structures
differently; there were a number of reasonable options.  For our purposes
here, a variant of Tarski's notion is more appropriate.  Respecting
tradition, we do not redefine structures.  Rather, we modify the notion of
structure and give the new notion a new name.

Structures without relations are called {\em algebras\/} in the branch of
mathematics called universal algebra.  Restrict attention to algebras with
distinct nullary operations {\true} and {\false} and define basic relations
as basic operations taking only the Boolean values {\true} and {\false}.
Further restrict attention to algebras with the equality relation and the
usual Boolean operations.  (We will specify later the values of the Boolean
operations outside their natural domains.)  The resulting notion of algebra
is our variant of the notion of structure with equality.  It allows us to
write quantifier-free formulas as terms.

Actually, we are interested in multi-sorted structures with partial
operations.  The sorts can be given by unary relations (they will be called
universes and the whole underlying set of a structure will be called the
superuniverse).  To deal with partial functions, further restrict attention
to algebras with a nullary operation {\undef}, different from {\true} and
{\em false}, and interpret an operation $f$ as undefined at a tuple $\bara$
if $f(\bara)=\undef$.  These algebras will be called {\em static
algebras\/} or {\em states\/}.  Their operations will be called functions.

In the following subsections, we start anew and define static algebras from
scratch, establishing terminology on the way.

\subsection{Vocabularies}

A {\em vocabulary\/} (or {\em signature\/}) is a finite collection of
function names, each of a fixed arity.  Some function names may be marked
as {\em relation names\/} or {\em static names\/}, or both.  Every
vocabulary contains the following static names: the equality sign, nullary
function names {\true}, {\false}, {\undef} and the names of the usual
Boolean operations.  The equality sign and {\true}, {\false} are marked as
relation names.  The Greek letter $\Ups$ is reserved to denote
vocabularies.

\paragraph{Logic Names}
The particular function names listed above are {\em basic logic names\/}.
There are precedents of logic names in mathematical logic, though usually
they are called logical constants.  For example, the equality sign is a
logic name in first-order logic with equality.  Usually, logic names are
present in every vocabulary and their interpretations satisfy some {\em a
priori\/} restrictions.  Accordingly, we suppose that the basic logic names
appear in every vocabulary, and thus there is no need to mention them when
a particular vocabulary is described.

An additional logic name is introduced in \S3.  It does not necessarily
appear in every vocabulary and it is not marked static.  The latter is one
reason why we do not use the term ``logical constants''.

\subsection{Definition of Static Algebras}

A {\em static algebra\/} or (for the sake of brevity) {\em state\/} $S$ of
vocabulary $\Ups$ is a nonempty set $X$, the {\em superuniverse\/} of $S$,
together with interpretations of the function names in $\Ups$ on $X$.  An
$r$-ary function name is interpreted as a function from $X^r$ to $X$, a
{\em basic function\/} of $S$.  The interpretation of an $r$-ary relation
name is a function from $X^r$ to $\{\true,\false\}$, a {\em basic
relation\/} of $S$.  The vocabulary $\Ups$ is called the {\em vocabulary\/}
of $S$ and denoted $\Fun(S)$.

The interpretations of the nullary logic names {\true}, {\false} and
{\undef} are distinct elements of $X$.  The Boolean operations behave in
the usual way on the Boolean values {\true} and {\false} and produce
{\undef} if at least one of the arguments is not Boolean.  The equality
sign is interpreted as the characteristic function of the identity relation
on $X$.  If $f(\barx)$ evaluates to {\true} in $S$, we say that $f(\barx)$
holds in $S$; and if $f(\barx)$ evaluates to $\false$ in $S$, we say that
$f(\barx)$ fails in $S$.

Formally speaking, basic functions are total.  However, we view them as
being partial and define the domain $\Dom(f)$ of an $r$-ary basic function
$f$ as the set of $r$-tuples $\barx$ such that $f(\barx)\neq\undef$.  Let
us stress though that {\undef} is an ordinary element of the superuniverse.
Often, a basic function produces {\undef} if at least one argument equals
{\undef}, but this is not required and there are exceptions ({\em e.g.\/}
basic relations).

\paragraph{Universes}
A basic relation $f$ may be viewed as the set of tuples where it evaluates
to {\true}.  We may write $\barx\in f$ instead of $f(\barx)$.  If $f$ is
unary it can be viewed as a special {\em universe\/}.  For example, we may
have a universe Nodes and declare a binary relation Edge over the universe
of Nodes; Edge$(x,y)$ will hold only if both $x$ and $y$ belong to Nodes.
Such universes allow us to view states as many-sorted structures.
Sometimes we speak about universe names.  These are unary relation names
intended to be used as universes.

As a rule, {\undef} is not included in universes.  Coming back to our
example, is it natural that Edge(\undef,\undef) equals {\false} rather than
{\undef}?  In a sense, yes.  Think about Edge as a set of pairs of nodes.
It is natural that the pair (\undef,\undef) does not belong there.

\subsection{Terms}

{\em Terms\/} are defined recursively, as in first-order logic:
\begin{itemize}\item
  A variable is a term.
\item If $f$ is an $r$-ary function name and $t_1,\ldots,t_r$ are terms,
  then $f(t_1,\ldots,t_r)$ is a term.
\end{itemize}

As usual, {\em ground terms\/} are terms without variables.  By analogy,
other syntactical objects without variables will be called ground.

{\em Atomic Boolean terms\/} are terms of the form $f(\bart\,)$, where $f$
is a relation name.  {\em Boolean terms\/} are built from atomic Boolean
terms by means of the Boolean operations.

\paragraph{Appropriate States and the Fun Notation}
In addition to terms, we will define various other syntactic objects, {\em
  e.g.\/}, update instructions and transition rules.  We call a state $S$
{\em appropriate\/} for a syntactic object $s$ if $\Fun(S)$ includes the
collection of function names that occur in $s$.  By default (that is,
unless explicitly defined differently), that collection will be denoted
$\Fun(s)$.

In an appropriate state $S$, a ground term $t=f(t_1,\ldots,t_r)$ evaluates
to an element \( Val_S(t) = f(Val_S(t_1),\ldots,Val_S(t_r)) \).  If
$\bart\,$ is a tuple $(t_1,\ldots,t_r)$ of terms, define \( Val_S(\bart\,)
= (Val_S(t_1),\ldots,Val_S(t_r)) \).

An expression $t_1=t_2$ may be a Boolean term or a metalanguage statement.
Often it does not matter which it is.  One can use two different equality
signs or just try to be careful; we choose the second alternative.

\subsection{Locations and Updates}

As in first-order logic, the {\em reduct\/} of an $\Ups$-state $S$ to a
smaller vocabulary $\Ups'$ is the $\Ups'$-state $S'$ obtained from $S$ by
``disinterpreting'' function names in $\Ups-\Ups'$; $S$ is an {\em
expansion\/} of $S'$ to $\Ups$.

A {\em carrier\/} is a state whose vocabulary contains only static function
names.  The {\em carrier $|S|$ of a state $S$\/} is the reduct of $S$ to
the static part of $\Fun(S)$.    

A {\em location over a carrier $C$\/} is a pair $\ell=(f,\barx)$, where $f$
is a function name outside of $\Fun(C)$ and $\barx$ is a tuple of elements
of $C$ whose length equals the arity of $f$; location $\ell$ is {\em
relational\/} if $f$ is a relation symbol.  $\Loc_\Ups(C)$ is the
collection of all locations over $C$ with function names in $\Ups$.  An
$\Ups$-state $S$ with carrier $C$ will sometimes be viewed as a function
from $\Loc_\Ups(C)$ to (the superuniverse of) $C$; {\em locations of $S$\/}
are locations in $\Loc_\Ups(C)$.

\vbox{If a state $S$ is appropriate for a ground term $t_0=f(\bart\,)$,
then the {\em location\/} of $t_0$ in $S$ is the location
$(f,Val_S(\bart\,))$.}

An {\em update\/} of a state $S$ is a pair $\al=(\ell,y)$, where $\ell$ is
a location of $S$ and $y\in|S|$; if $\ell$ is relational then $y$ is
Boolean.  (More precisely, $y$ belongs to the superuniverse of static
algebra $|S|$; the looser language is common in logic.)  The location
$\ell$ is the {\em location\/} $\Loc(\al)$ of $\al$, and $y$ is the {\em
value\/} $\Val(\al)$ of $\al$.  To {\em fire\/} $\al$ at $S$, put $y$ into
the location $\ell$; that is, redefine $S$ to map $\ell$ to $y$.  The
result is a new state $S'$ such that $\Fun(S')=\Fun(S)$, $|S'|=|S|$,
$S'(\ell)=y$ and $S'(\ell')=S(\ell')$ for every location $\ell'$ of $S$
different from $\ell$.

\subsection{Update Sets and Families of Update Sets}

An {\em update set\/} $\be$ over a state $S$ is a set of updates of $S$.
$\Loc(\be)=\{\Loc(\al)\st \al\in\be\}$.  For each $\ell\in\Loc(\be)$,
\( \Val_\be(\ell)  = \{ \Val(\al) \st \al\in\be \w \Loc(\al)=\ell\}\).

An update set $\be$ is {\em consistent\/} at the given state $S$ if every
$\Val_\be(\ell)$ is a singleton set; otherwise $\be$ is inconsistent.

To fire a consistent $\be$ at the given state $S$, fire all its members
simultaneously.  The result is a new state $S'$ with the same vocabulary
and carrier as $S$.  If $\ell\in\Loc(\be)$ then $S'(\ell)$ is the only
element of $\Val_\be(\ell)$; otherwise $S'(\ell)=S(\ell)$.  To fire an
inconsistent update set $\be$ at the given state $S$, do nothing; the new
state $S'$ equals $S$.

\subparagraph{Remark} It is reasonable to require that the detection of
inconsistency manifest itself in some way; for example, a nullary function
{\em crash\/} automatically gets value {\em true}.  To keep the EA logic
clean and simple, we try to minimize the number of things done
automatically, and thus we leave necessary manifestations of inconsistency
to the programmer.  This is one application of the pragmatic Occam's razor
of \S1; substantial programming convenience has not been demonstrated
yet.\ss

To fire a family $\ga$ of update sets over $S$, nondeterministically choose
some update set $\be\in\ga$ and fire it at $S$.  If $\ga=\emptyset$, do
nothing.  Intentionally, the empty family of update sets means
inconsistency.

\subsection{Conservative Determinism vs.\ Local Nondeterminism}

The mode of dealing with inconsistent update sets described above can be
called conservative determinism.  The mode of dealing with inconsistent
update sets in the tutorial was different: Fire all updates simultaneously;
in case of conflict at any location $\ell$, choose the new value for $\ell$
nondeterministically among all candidate values.  It could be called local
nondeterminism.

With the exception of this change in the treatment of inconsistent update
sets, this guide is compatible with the tutorial.  The change is not as big
as it may seem because people are usually interested in deterministic
programs.  As far as we know, no existing EA application is affected.  The
local nondeterminism has not been exploited.  The conservative determinism
is simpler, and a more manageable form of nondeterminism will be introduced
in \S4.

\section{Sequential Evolving Algebras}

Basic transition rules are defined in subsection~3.1.  Subsection~3.2 deals
with the problem of extending universes.  The reader may skip 3.2 and go
directly to subsection~3.3 on programs and runs.

\subsection{Basic Transition Rules}

In this subsection, terms are ground.

\subsubsection{Update Instructions}

An {\em update instruction\/} $R$ is an expression

\[ f(\bart\,) := t_0 \]

\ni where $f$ is a non-static function name (the {\em subject\/} of the
instruction), $\bart\,$\/ is a tuple of terms whose length equals the arity
of $f$, and $t_0$ is another term; if $f$ is a relation name then $t_0$
must be a Boolean term.  (Update instructions are called local function
updates in the tutorial.)

\paragraph{Semantics}
To execute $R$ at an appropriate state $S$, fire the update $\al=(\ell,y)$
at $S$, where $\ell=(f,Val_S(\bart\,))$ and $y=Val_S(t_0)$.  For future
reference define $\Updates(R,S) = \{\al\}$.

\subsubsection{Two Rule Constructors}

{\em Basic rules\/} are constructed recursively from update instructions by
means of two rule constructors: the sequence constructor and the
conditional constructor.  Semantics is defined by means of update sets.
For each rule $R$ and every state $S$ appropriate for $R$, we define an
update set $\Updates(R,S)$ over $S$.  To fire $R$ at $S$, fire
$\Updates(R,S)$.

\paragraph{The Sequence Constructor}  
A sequence of rules is a rule.

\subparagraph{Semantics}
If $R$ is a sequence of rules $R_1,\ldots,R_k$ then
\[ \Updates(R,S) = \Updates(R_1,S)\cup\cdots\cup\Updates(R_k,S). \]
In other words, to fire a sequence of rules, fire all of them
simultaneously.  Notice that $\Updates(R,S)$ is inconsistent if any $R_i$
is so.

\subparagraph{Remark} 
The term ``sequence'' may be misleading here.  We are not executing first
$R_1$, then $R_2$, then $R_3$, {\em etc.\/}.  A better term is ``block''.
(This remark is written at the proofreading stage.)

\paragraph{The Conditional Constructor}
If $k$ is a natural number, $g_0,\ldots,g_k$ are Boolean terms and
$R_0,\ldots,R_k$ are rules, then the following expression is a rule:

\ni\parbox{4in}{
\begin{eatab}
if $g_0$ then $R_0$\\ 
elseif $g_1$ then $R_1$\\
\> $\vdots$\\
elseif $g_k$ then $R_k$\\ 
endif
\end{eatab}}
 
If the guard $g_k$ is the nullary function {\true}, then the last elseif
clause may be replaced by ``else $R_k$''.  For brevity we will say that the
conditional rule $R$ above is the conditional rule with clauses
$(g_0,R_0),\ldots,(g_k,R_k)$.

\subparagraph{Semantics}
$\Updates(R,S)=\Updates(R_i,S)$ if $g_i$ holds in $S$ but every $g_j$ with
$j<i$ fails in $S$.  $\Updates(R,S)=\emptyset$ if every $g_i$ fails in $S$.

\subsubsection{Guarded Multi-updates}

A {\em multi-update instruction\/} is a sequence of update instructions.  A
{\em guarded update instruction\/} (respectively, {\em guarded multi-update
instruction\/}) is a rule of the form

\begin{eatab}
if $g$ then $R$ endif
\end{eatab}

\ni where $R$ is an update (respectively, a multi-update) instruction.

\begin{Lemma}
For every rule $R$, there is a sequence $R'$ of guarded updates such that
$\Fun\,(R')=\Fun\,(R)$ and $\Updates\,(R',S)=\Updates\,(R,S)$ for all
appropriate states $S$.
\end{Lemma}

For example, the rule

\ni\parbox{4in}{
\begin{eatab}
 if FirstChild(c)$\neq\undef$ then c:=FirstChild(c)\\
 elseif NextSib(c)$\neq\undef$ then c:=NextSib(c)\\
 elseif Parent(c)$\neq\undef$ then c:=Parent(c)\\
 endif
\end{eatab}}

\ni converts to the following sequence of guarded updates:

\ni\parbox{4in}{
\begin{eatab}
 if FirstChild(c)$\neq\undef$ then c:=FirstChild(c) endif\\
 if FirstChild(c)=\undef\/ and NextSib(c)$\neq\undef$ then\\
 \> c:=NextSib(c) endif\\
 if FirstChild(c)=\undef\/ and NextSib(c)=\undef\\
 \> and Parent(c)$\neq\undef$ then c:=Parent(c) endif
\end{eatab}}

The Lemma suggests a simpler definition of rules.  The reason for choosing
the recursive definition is pragmatic.  It is too tedious to write rules as
sequences of guarded updates.  It is feasible to write them as sequences of
guarded multi-updates but it is more convenient and practical to use elseif
clauses and nest conditionals.  The pragmatic Occam's razor does not cut as
much as the original Occam's razor would.

\subparagraph{Remark}
This is another proofreading time remark.  The new version of the EA
interpreter permits the use of two additional rule constructors.  One is
the case constructor, like that in Pascal, which may make the execution
substantially more efficient.  Of course, the same set of updates is
generated by a case command and its case-free equivalent; the difference is
in how fast this set is generated.  For example, consider a sequence of
rules of the form ``if $t=i$ then $R_i$ endif'' where $i$ ranges from $1$
to a relatively large $n$.  This example is extreme, because the the set
$\{1,\ldots,n\}$ of alternatives is so easy to deal with; but it is not
unusual to have similar long sequences of rules.  In addition, the case
construct makes it easier to program a sequential execution of a sequence
of rules, which is sometimes desirable.  The other rule constructor is
``let x=t in R'', which prevents re-evaluations of term $t$ in $R$ and
which has been used informally.  The let constructor was advocated by Raghu
Mani who is working on the new EA interpreter.

\subsection{Importing New Elements}

The basic rules suffice for many purposes ({\em e.g.\/}, for describing the
C programming language \cite{GH}), but they do not suffice to model all
sequential algorithms.  A sequential algorithm may add a new node to a
graph or create a new message.  We need rules that allow us to create new
nodes, new messages, {\em etc.\/}, and such rules are introduced in this
subsection.  However, we do not create new elements; instead, we use a
special universe Reserve from which the new elements come.

In this section we use individual variables, but only in a limited way.
(Variables are used more extensively in \S5.)  Roughly speaking, only bound
variables are used; free variables appear only in contexts where some
values have been assigned to them.

\subsubsection{Reserve}

In addition to basic logic names, we introduce a new logic name: a universe
name Reserve.  It is not static, and we do not require that it belong to
the vocabulary of every static algebra.  If the vocabulary of state $S$
contains Reserve, then the set $\{x\st S\models x\in\Reserve\}$ is the {\em
reserve\/} of $S$.  Intuitively the reserve is a naked set.

\paragraph{Reserve Proviso}
Every state satisfies the following conditions:
\begin{itemize}\item
Every basic relation, with the exception of equality and Reserve, evaluates
to {\false} if at least one of its arguments belongs to the reserve.
\item
Every other basic function evaluates to {\em undef\/} if at least one of
its arguments belongs to the reserve.
\item
No basic function outputs an element of the reserve.
\end{itemize}

It follows that every permutation of the reserve is an automorphism of the
state.

\subsubsection{Transition Rules:\ \ Syntax}

Generalize the definitions of terms and update instructions in 3.1 as
follows:
\begin{itemize}

\item allow terms to have variables, and

\item forbid mentioning Reserve.
\end{itemize}

Variables are often treated as auxiliary nullary function names below but
{\em a variable cannot be the subject of an update instruction}.  The
reason for forbidding to mention Reserve in terms and update instructions
is discussed below.

{\em Rules\/} are constructed from update instructions by means of three
rule constructors: the sequence constructor, the conditional constructor
and the import constructor.

\paragraph{The Import Constructor}  
If $v$ is a variable and $R_0$ is a rule, then the following expression is
a rule with {\em main existential variable\/} $v$ and {\em body\/} $R_0$\/:

\ni\parbox{4in}{
\begin{eatab}
import $v$\\
\> $R_0$\\
endimport
\end{eatab}}

In the usual and obvious way define which occurrences of variables are free
and which are bound. Call a rule {\em perspicuous\/} if no variable has
both bound and free occurrences, and no bound variable is declared more
than once.  (The latter means here that different occurrences of the import
command have different main existential variables.)

Let $\Free(R)$ be the set of free variables of a rule $R$.  In other words,
$\Free(R)$ is the set of variables $v$ such that $v$ occurs freely in rule
$R$.  Define $\Bound(R)$ similarly.  If $R$ is an import rule with main
existential variable $v$ and body $R_0$, we have:
\[ \Free(R)=\Free(R_0)-\{v\}, \mbox{\ \ and\ \ }
   \Bound(R)=\Bound(R_0) \cup \{v\}. \]

\subsubsection{Auxiliary Vocabularies}

The names of variables are different from function names of course, but it
is convenient to treat free variables of rules as auxiliary nullary
functions (which cannot be subjects of update instructions).  An {\em
auxiliary vocabulary\/} has the form $\Ups\cup V$, where $\Ups$ is a
genuine vocabulary and $V$ is a finite set of variables.  

If $S$ is a state of an auxiliary vocabulary $\Ups' =\Ups\cup V$, then
$\Fun(S) =\Ups'$.  $S$ is {\em appropriate\/} for a rule $R$ if $\Ups$
contains all function names of $R$ and $V$ contains all free variables of
$R$.  $R$ is {\em $S$-perspicuous\/} if it is perspicuous and its bound
variables do not occur in $V$.

\subsubsection{Transition Rules:\ \ Semantics}

An import commands chooses an element of the reserve and removes it from
the reserve.  To clarify our intentions, we note that the non-perspicuous
rule

\ni\parbox{4in}{
\begin{eatab}
import $v$\\
\> Parent($v$):=CurrentNode\\
endimport\\
import $v$\\
\> Parent($v$):=CurrentNode\\
endimport
\end{eatab}}

\ni creates {\em two\/} children of CurrentNode.  In general, different
choices from the reserve produce different elements.

For each rule $R$ and every state $S$ appropriate for $R$, we define an
update set $\Updates(R,S)$ over $S$; to fire $R$ at $S$, fire
$\Updates(R,S)$.

First, we consider the case of when $R$ is $S$-perspicuous.  Fix an
injective map $\xi$ from $\Bound(R)$ to the reserve of the given $S$.  (The
injectivity means that $\xi$ assigns different elements to different bound
variables.)  By induction on subrule $R'$ of $R$ we define sets
$\Updates(R',S',\xi)$ where $S'$ is an expansion of $S$ appropriate for
$R'$ and such that $R'$ is $S'$-perspicuous.  (Recall that $S'$ is an
expansion of $S$ if and only if the reduct of $S'$ to $\Fun(S)$ equals
$S$.)  Let $\Ups' =\Fun(S')$.

The cases of update instructions, sequence rules and conditional rules are
treated as above.  (Variables in $\Ups'$ are treated as nullary functions.)
Suppose that $R'$ is an import rule with main existential variable $v$ and
body $R_0$.  Let $a=\xi(v)$ and $S_a'$ be the expansion of $S'$ to the
auxiliary vocabulary $\Ups'\cup\{v\}$ where $v$ is interpreted as $a$.
Recall that variables are not subjects of update instructions.  Thus
$\Updates(R_0,S_a',\xi)$ is an update set over $S'$.  Set
\[ \Updates(R',S',\xi) = 
\{ ((\Reserve,a),\false) \} \cup \Updates(R_0,S_a',\xi). \]

Finally $\Updates(R,S) =\Updates(R,S,\xi)$.  Of course, $\Updates(R,S)$ is
not defined uniquely, because it depends on $\xi$.  It is easy to see,
however, the resulting state is unique up to isomorphism.

Second, we stipulate that an arbitrary rule $R$ is equivalent, over the
given appropriate state $S$, to an $S$-perspicuous rule $R'$ obtained from
$R$ by renaming the bound variables.  (The desired $R'$ can be obtained by
iterating the following transformation: Select an innermost import subrule
$R_1$ whose main existential variable $v$ occurs in the rest of the rule or
in $\Fun(S)$, and replace $v$ with a fresh variable in $R_1$.)  The
stipulation means the following: To fire $R$ at $S$, fire $R'$ at $S$.

\paragraph{Discarding Elements from Universes}
Finally, we explain the reason for forbidding to mention Reserve explicitly
in our rules.  Terms $\Reserve(t)$ always evaluate to $\false$, so
evaluating $\Reserve(t)$ or setting it to $\false$ is useless.  But why not
to allow putting the value {\true} into Reserve locations.  Elements can be
discarded from universes, of course; to discard an element (represented by
a term) $t$ from a universe $U$, use the instruction $U(t):=\false$.  Isn't
the reserve a natural place for unwanted elements?  Yes, it is.  Notice,
however, that moving an element into the reserve may necessitate numerous
changes of basic functions in order to ensure that the Reserve proviso
remains valid.  Would such a move contradict the sequential character of
our rules?  Not necessarily.  We could just mark discarded elements as
reserve elements, but then it might be necessary to augment rules with
numerous guards $\Reserve(t)=\false$, which would be too tedious.  It is
preferable to leave the discarded elements alone.  This pragmatic argument
was put forward originally by Egon B\"orger.

But shouldn't the computational resources of the ealgebra simulating an
algorithm $A$ closely reflect the computational resources of $A$?  Yes, but
it is important to separate the following concerns: the logic of $A$ and
the relevant resources of $A$.  Concentrating on the logic of $A$ may
allow one to come up with simpler rules for the simulating ealgebra.  And
if one needs to track the resources of $A$, a separate bookkeeping may be
set up.  This separation of concerns allows us, for example, to use
infinite universes.  And caring about only particular elements and
universes, rather than the whole superuniverse, makes combining ealgebras
easier.

\subsubsection{Importing Several Elements at a Time}

Let $v_1,v_2$ be distinct variables.  Abbreviate

\ni\parbox{1.25in}{
\begin{eatab}
import $v_1$ \\
\> import $v_2$ \\
\> \> $R_0$\\
\> endimport\\
endimport
\end{eatab}}
\hfill to \hfill
\parbox{1.75in}{
\begin{eatab}
import $v_1,v_2$ \\
\> $R_0$\\
endimport
\end{eatab}}

In a similar way, define abbreviations

\ni\parbox{4in}{
\begin{eatab}
import $v_1,\ldots,v_k$ \\
\> $R_0$\\
endimport
\end{eatab}}

Abbreviate

\ni\parbox{1.25in}{
\begin{eatab}
import $v_1,\ldots,v_k$ \\
\> $U(v_1):=\true$\\
\> $\vdots$\\
\> $U(v_k):=\true$\\
\> $R_0$\\
endimport
\end{eatab}} 
\hfill to \hfill
\parbox{1.75in}{
\begin{eatab}
  extend $U$ with $v_1,\ldots,v_k$ \\ \> $R_0$\\ endextend
\end{eatab}}

Later (in 5.4) we'll see how to import a number of elements that is not
bounded {\em a priori\/} by any constant.  Here is an example of the extend
rule:

\ni\parbox{4in}{
\begin{eatab}
extend Nodes with $v_1,v_2$\\
\> FistChild(CurrentNode) := $v_1$\\
\> SecondChild(CurrentNode) := $v_2$\\
\> NextSib($v_1$) := $v_2$\\
endextend
\end{eatab}}

\subsection{Programs and Runs}

\subsubsection{Programs and Pure Runs}

A {\em program\/} $P$ is a rule without free variables.  A {\em basic
program\/} is a basic rule without free variables.  In applications, a
program is usually a sequence of rules referred to as rules of the program.
To fire $P$ at an appropriate state $S$, fire $\Updates(P,S)$ at $S$.

A {\em pure run\/} of $P$ is a sequence
\( \langle S_n \st n<\kappa \rangle \)
of states of vocabulary $\Fun(P)$ such that each $S_{n+1}$ is obtained from
$S_n$ by firing $P$ at $S_n$.  Here and henceforth $\kappa$ is a positive
integer or the first infinite ordinal.  In the latter case, $\{n\st n<\kappa\}$
is the set of all natural numbers.

The adjective ``pure'' reflects the fact that the run is not affected by
the environment.

\subsubsection{External Functions}

In general runs may be affected by the environment.  Suppose that the
environment manifests itself via some basic functions $e_1,\ldots,e_k$,
called {\em external functions\/}.  A typical external function is the
input provided by the user.

Think about an external function as a (dynamic) oracle.  The ealgebra
provides the arguments and the oracle gives the result.  The oracle need
not be consistent and may give different results for the same argument at
different times.  The seeming inconsistency may be quite natural.  For
example, the argument may specify an input channel.  The next time around,
another input can come via the same channel.

However, the oracle should be consistent during the execution of any one
step of the program.  In an implementation, this may be achieved by not
reiterating the same question during a one-step execution.  Ask the
question once and, if necessary, save the result and reuse it.

The computation steps of a program are supposed to be atomic at an
appropriate level of abstraction.  A computation step is hardly atomic if
during that step the ealgebra queries an oracle and then, depending on the
result, submits another query to the same or a different oracle.  Thus it
seems reasonable to forbid nesting of external functions.  Indeed, the need
to nest external functions has not arisen in applications so far.  But we
withhold final judgement and wait for more experimentation.

Call non-external basic functions {\em internal\/}.  If $S$ is an
appropriate state for a program $P$, let $S^-$ be the reduct of $S$ to the
internal vocabulary.

\paragraph{Runs}
A {\em run\/} of a program $P$ is a sequence
\( \langle S_n \st n<\kappa \rangle \)
of states where:
\begin{itemize}\item
every nonfinal $S_n$ is an appropriate state for $P$ and the final state
(if any) is a state of the internal vocabulary of $P$, and
\item
every $S_{n+1}^-$ is obtained from $S_n$ by firing $P$ at $S_n$.
\end{itemize}

\paragraph{Internal and External Locations}
It may happen that the environment controls only a part of a function $e_i$
and the remaining part of $e_i$ is governed internally.  In such a case it
is natural to speak about internal and external locations rather than
internal and external functions.  See an example in \cite[3.1]{BGR}.  The
generalization to that case is relatively straightforward.

\paragraph{Irrelevant Values of External Functions}
In order to fire a given program at a given state, we may not need to know
all about the state.  Only some values of external functions may be needed
for firing.  We may not care about or even know the values of external
functions which are not needed for the execution.  Some of those values may
even be ill-defined.  There is also an issue of influencing the environment
by requiring an extra value, {\em e.g.\/}, by requiring a user-provided
datum.

It is natural to set all irrelevant values of external functions to
{\undef}.  However, caution should be exercised in the distributed
situation (see \S6) where other agents may have different views of those
values.

Sometimes it may be simpler to use partial states.  A partial $\Ups$-state
$S$ with carrier $C$ can be defined as a partial function from
$\Loc_\Ups(C)$ to $C$.  See examples in \cite{BGR,GM}.  For simplicity, we will
not use partial states here.

\section{Nondeterministic Sequential Ealgebras and Some Other
Simple Extensions of the Basic Model} 

Describing algorithms on higher abstraction levels, one often comes across
the phenomenon of nondeterminism.  Nevertheless, the built-in
nondeterminism of ealgebras has been rarely used.  It is often more
appropriate to use external functions to reflect nondeterministic behavior.
(In the distributed case, nondeterminism may be often eliminated by
introducing additional agents.)  Consider for instance the assignment
statement of the C programming language.  Should one evaluate the left side
or the right side first?  According to the ANSI standard (ANSI is the
American National Standards Institute), the choice of the evaluation order
is implementation-dependent.  Moreover, an implementation does not have to
be consistent; the evaluation order may change when the same assignment
statement is executed next time around (say, in a loop).  This is an
obvious case of nondeterminism and first we, the authors of \cite{GH}, were
tempted to use a nondeterministic rule to reflect the nondeterminism.  But
then we realized that C is perfectly deterministic.  It is just that
execution may depend on information provided by implementation.  Thus it is
more faithful to the standard (and more convenient) to use an external
function that decides the evaluation order.

Still, nondeterministic commands may be desired and we provide such
commands in this section.  For example, it may be convenient to formalize
the environment in a distributed situation, so that an external function of
one agent is nondeterministically computed by another agent.

For simplicity, we ignore the import constructor in this section.  It is
easy to extend the language of this section with the import constructor.
Moreover, the choice constructor defined below and the import constructor
can be combined into one constructor. 

\subsection{Basic Evolving Algebras with Choice}

\subsubsection{Syntax}

Transition rules are constructed as in 3.2, except that instead of the
import constructor, we use the Choose (or Choice) Constructor:

\paragraph{Choose Constructor}
If $U$ is a universe name different from Reserve, $v$ is a variable and
$R_0$ is a rule then the following expression is a rule with {\em main
existential variable\/} $v$ that ranges over $U$ and {\em body\/} $R_0$\/:

\ni\parbox{4in}{
\begin{eatab}
choose $v$ in $U$\\
\> $R_0$\\
endchoose
\end{eatab}}

This is the basic version of the choice constructor; a stronger version is
defined in 4.2.2.  Perspicuity is defined as 3.2.

\subsubsection{Semantics}

For each rule $R$ and each state $S$ appropriate for $R$, we define
a family $\ga=\NUpdates(R,S)$ of update sets over $S$.  To fire $R$ at $S$,
choose any $\be\in\ga$ and fire $\be$ at $S$. 

We stipulate that an arbitrary rule $R$ is equivalent, over the given $S$,
to an $S$-perspicuous rule $R'$ obtained from $R$ by renaming the bound
variables.  The equivalence means here that $\NUpdates(R,S)
=\NUpdates(R',S)$.  It remains to define $\ga=\NUpdates(R,S)$ when $S$ is
$S$-perspicuous.

\paragraph{Global Choice Semantics}

Semantics is defined as in 3.2.4.  On one hand, things are simpler this
time around because there is no correlation among individual choices.  On
the other hand, there is a complication related to attempts to choose an
element of the empty set.  Such attempt cannot succeed and the execution
should be aborted.  To deal with this complication, we extend the
collection of updates of any state by an ideal element $\bot$ that
symbolizes inconsistency.  If an update set $\be$ contains $\bot$ then
firing $\be$ does not change the state; we call such $\be$ contradictory.

Suppose that a state $S$ is appropriate for a rule $R$ and $R$ is
$S$-perspicuous.  Let $V$ be the collection of bound variables of $R$ such
that the range of $v$ is not empty in state $S$.  Fix a function $\xi$ on
$V$ such that, for each $v\in V$, $\xi(v)$ belongs to the range of $v$ in
$S$.  By induction on subrule $R'$ of $R$ define $\Updates(R',S',\xi)$
where $S'$ is an expansion of $S$ appropriate for $R'$ and $R'$ is
$S'$-perspicuous.

The cases of update instructions, sequence rules and conditional rule are
treated as above.  Notice that if $R'$ is a sequence of rules $R_i$ and
some $\Updates(R_i,S',\xi)$ is contradictory then $\Updates(R',S',\xi)$ is
so.

Suppose that $R'$ is a choose rule with main existential variable $v$ and
body $R_0$.  If the range of $v$ is empty then $\Updates(R',S',\xi) =\bot$.
Otherwise let $a=\xi(v)$ and $S_a'$ be the expansion of $S'$ to the
auxiliary vocabulary $\Ups'\cup\{v\}$ where $v$ is interpreted as $a$.  Set
$\Updates(R',S',\xi) =\Updates(R_0,S_a',\xi)$.

Finally, $\NUpdates(R,S)$ is the set of $\Updates(R,S,\xi)$ where $\xi$
takes all possible values.

\paragraph{Semantics without Global Choice}

The global choice semantics is straightforward.  However, contrary to the
situation 3.2.4, there is no correlation among individual choices this time
around, and thus there is no real need for a global choice function $\xi$.
It may be more elegant to define $\ga=\NUpdates(R,S)$ directly by induction
on $R$.  We suppose again that $S$ is appropriate to $R$ and $R$ is
$S$-conspicuous.

If $R$ is an update instruction then $\ga =\{\Updates(R,S)\}$.  If $R$ is a
sequence of rules $R_1,\ldots,R_k$, then
\[ \ga = \{\be_1\cup\cdots\cup \be_k\st
\mbox{ each } \be_i\in\NUpdates(R_i,S)\}.\]
Notice that $\ga$ is empty if some so is $\NUpdates(R_i,S)$.

If $R$ is a conditional rule with clauses $(g_0,R_0),\ldots,(g_k,R_k)$, we
have two cases as usual; if all $k+1$ guards fail in $S$ then $\ga
=\{\emptyset\}$, and if $g_i$ is the first guard that holds in $S$ then
$\ga =\NUpdates(R_i,S)$.  (It would be a mistake to replace $\{\emptyset\}$
with $\emptyset$ above.  If $\NUpdates(R_1,S) =\emptyset$ then
$\NUpdates((R_1,R_2),S) =\emptyset$ for every rule $R_2$, which is not
desired.)

Finally, suppose that $R$ is a choose rule with universe name $U$, main
existential variable $v$ and body $R_0$.  For each $a\in U$, let $S_a$ be
the expansion of $S$ of the auxiliary vocabulary $\Fun(S)\cup\{v\}$ where
$v$ is interpreted as $a$.  Then
\[ \ga = \bigcup \{\NUpdates(R_0,S_a)\st a\in U\}. \]

\ni Notice that $\ga$ is empty if $U$ is empty.  

It is easy to check that if $R$ contains no choice subrules then
$\NUpdates(R,S)=\{\Updates(R,S)\}$.

\paragraph{Remark}  
In the second approach, $\bot$ is not used.  Its role is played by the
empty family of update sets.  This gives us an idea to eliminate the use of
$\bot$ in the first approach: replace $\Updates(R',S',\xi)$ with the
singleton family $\{\Updates(R',S',\xi)\}$ and replace $\bot$ with the
empty family.

\paragraph{Runs}

The definition of runs in \S3 remains in force.

\subsubsection{Abbreviations}

Let $v_1,v_2$ be distinct variables.  Abbreviate

\ni\parbox{1.5in}{
\begin{eatab}
choose $v_1$ in $U$\\
\> choose $v_2$ in $U$\\
\> \> $R_0$\\
\> endchoose\\
endchoose
\end{eatab}}
\hfill to \hfill
\parbox{1.5in}{
\begin{eatab}
choose $v_1,v_2$ in $U$\\
\> $R_0$\\
endchoose
\end{eatab}}

In a similar way define abbreviation

\ni\parbox{4in}{
\begin{eatab}
choose $v_1,\ldots,v_k$ in $U$\\
\> $R_0$\\
endchoose
\end{eatab}}

\subsection{Some Other Simple Extensions of the Basic Model}

We consider three extensions, which are simple in the sense that it is easy
to define them.  The third extension has not been used; it is just a trial
balloon.

\subsubsection{First-order Guards}

In \S3, guards were Boolean terms.  Now we introduce a separate syntactic
category of guards.  Intuitively, guards are first-order formulas with
bound variables.  It is intended that bound variables range over finite
domains, though exceptions are possible.  Here is a recursive definition:

\begin{itemize}\item
  If $f$ is an $r$-ary relation name and $t_1,\ldots,t_r$ are terms, then
  $f(t_1,\ldots,t_r)$ is a guard.
\item Any Boolean combination of guards is a guard.
\item If $g$ is a guard and $U$ a universe name, then $(\exists v\in U) g$
  and $(\forall v\in U) g$ are guards.
\end{itemize}

Call a guard closed if it has no free variables.  Extend the definition of
basic ealgebras by replacing the condition ``$g_1,\ldots,g_k$ are Boolean
terms'' with the condition ``$g_1,\ldots,g_k$ are closed guards'' in the
definition of the conditional rule constructor.

\subparagraph{Semantics}
The definition of the value of a closed guard at an appropriate state
mirrors the truth definition of formulas in first-order logic.  The
semantics of  rules is given exactly as in 3.1.

\paragraph{Remark}  
One can go further in this direction and use quantification inside other
terms.  To formalize this idea, the notion of terms can be redefined as
follows:
\begin{itemize}\item
  A variable $v$ is a term.
\item If $f$ is an $r$-ary function name and $t_1,\ldots,t_r$ are terms,
  then $f(t_1,\ldots,t_r)$ is a term.  The new term is Boolean if $f$ is a
  relation name.
\item Boolean terms are closed under the Boolean operations and
  quantification, and every Boolean term is a term.
\end{itemize}

\subsubsection{Qualified Choose Construct}

Restricting the choice by a Boolean term gives a much more powerful version
of the choose constructor.

\paragraph{Qualified Choose Constructor}
If $U$ is a universe name different from Reserve, $v$ is a variable, $g(v)$
is a Boolean term and $R_0$ is a rule, then the following expression is a
rule with {\em main existential variable\/} $v$ that ranges over $U$ and
body $R_0$\/:

\ni\parbox{4in}{
\begin{eatab}
  choose $v$ in $U$ satisfying $g(v)$\\ 
  \> $R_0$\\ 
  endchoose
\end{eatab}}

Replacing the choose constructor with the qualified choose constructor
requires only a small and obvious change in the semantical definition of
4.1.2.  We restrict attention to the global choice approach.  Consider the
case in the inductive definition of $\Updates(R',S',\xi)$ where $R'$ is a
choose rule and the range $U$ of the main existential variable $v$ of $R$
in $S'$ is not empty.  If $g(\xi(v))$ fails in $S$, set
$\Updates(R',S',\xi) =\bot$.

It is easy to construct a rule to choose several elements $v_1,\ldots,v_k$
subject to a condition $g(v_1,\ldots,v_k)$.

The qualified choose constructor may be too powerful.  The decision problem
whether there is any tuple $(v_1,\ldots,v_k)$ in the universe $U$
satisfying the condition $g$ may be hard.  If $U$ is the set of natural
numbers and $g$ a polynomial, the decision problem may even be undecidable
\cite{Mt}.  But the logical clarity of the constructor is attractive.  It may be
used in particular to reflect environmental forces that are not necessarily
algorithmic.

\subsubsection{Duplication}

The powerful extension of basic ealgebras considered in this subsection is
logically clear but untried and computationally expensive.  It does not
hurt to explore it though.

Call elements $a$ and $a'$ of a state $S$ indistinguishable as arguments
for a basic $r$-ary function $f$ if $f(b_1,\ldots,b_r) = f(c_1,\dots,c_r)$
for all $r$-tuples $b_1,\ldots,b_r$ and $c_1,\dots,c_r$ such that either
$b_i=c_i$ or $\{b_i,c_i\}=\{a,a'\}$.  Call $a,a'$ indistinguishable as
arguments if they are indistinguishable as arguments for any basic function
with the exception of equality.  Now we are ready to introduce the
duplicate constructor:

\ni\parbox{4in}{
\begin{eatab}
duplicate $t$ as $v$\\
\> $R_0$\\
endduplicate
\end{eatab}}

\subparagraph{Semantics} To execute, calculate $a=Val_S(t)$, get some $a'$
from the reserve and redefine basic functions on tuples involving $a'$ in
such a way that $a$ and $a'$ become indistinguishable as arguments.  Then
execute $R_0$ with $v$ equal $a'$.\ss

Duplication can be seen as a powerful inheritance mechanism.  It is easy to
see that the extend construct is not powerful enough to replace
duplication.

\section{Parallelism: Evolving Algebras with Variables}

What does it mean that an algorithm is sequential?  This usually means that
the algorithm has the following two features.  First, time is sequential.
The algorithm proceeds from some initial state $S_0$ to a state $S_1$, then
to a state $S_2$, {\em etc.\/}, and the steps are atomic.  Second, only a
bounded amount of work is done at each step.  In principle, a single agent
is able to move the algorithm from $S_0$ to $S_1$, then to $S_2$, {\em
etc.\/}.

In this section, we are interested in one-agent algorithms where the agent
may perform a substantial amount of work at one step.  We use variables to
formalize such algorithms.  It is intended that non-Reserve variables range
over finite (better yet, feasible) domains, though exceptions are possible.

We do not assume any particular sequential order of executing one step of
the algorithm.  It is possible that this work involves plenty of
parallelism and is implemented by a number of auxiliary agents.  But on the
natural level of abstraction of the given algorithm, those auxiliary agents
are invisible, and in principle a single agent may execute the algorithm.

\subsection{Variables}

In preceding sections, we dealt with implicit variables declarations by
means import commands, bounded quantifiers, etc.  In this section, we
introduce explicit variable declarations.

An {\em explicit atomic variable declaration\/} is an expression ``Var $v$
ranges over $U$'', where $v$ is a variable and $U$ a universe name.  The
universe $U$ is the range (or type) of the variable $v$.  A {\em explicit
variable declaration\/} $D$ is a sequence of explicit atomic variable
declarations, and $\Var(D)$ is the collection of variables in $D$.
For brevity, the adjective explicit is often omitted.

Intuitively, $D$ is a set of explicit atomic declarations, but we do not
forbid re-declarations of the same variable.  The range of a variable
$v\in\Var(D)$ is the range in the last declaration of $v$ in $D$.  In other
words, later declarations of a variable override the earlier ones.  One may
use more concise explicit variable declarations, like ``Var $v_1,\ldots,
v_k$ range over $U$''.

A variable declaration $D$ {\em covers\/} a syntactic object $s$ if
$\Var(D)$ contains all free (that is undeclared) variables of $s$.

As in 3.2.3, we use auxiliary vocabularies of the form $\Ups\cup V$, where
$\Ups$ is a genuine vocabulary, $V$ a finite set of variables and each
$v\in V$ is treated as a nullary function, except it cannot be the subject
of an update instruction.  We say that a state $S$ of an auxiliary
vocabulary is {\em appropriate\/} for a syntactical object $s$ if all
function names and all free variables of $s$ occur in $\Fun(S)$.

\subsection{Terms and Guards}

Terms and Boolean terms are defined in \S3.  Guards are defined in 4.2.1.
The free variables of terms and guards are defined inductively, as in
first-order logic.  Notice that a bounded quantifier implicitly contains an
atomic declaration.

As usual, every guard $g$ is equivalent to a guard $g'$ where no variable
is both bound and free and where different quantifier occurrences bind
different variables.  To reduce $g$ to $g'$, iterate the following
transformation: Select an innermost quantifier $q$ whose variable $v$
occurs outside the scope of $q$ and then replace $v$ with a fresh variable
in the scope of $q$.

\subsection{A Parallel Version of the Basic EA Model}

\subsubsection{Syntax} 

Update instructions and basic rules are defined as in 3.1, except that
terms may have free variables, and guards are defined as above.  In
addition, we have the following third rule constructor.

\paragraph{The Declaration Constructor} 
An atomic variable declaration followed by a rule is a rule.\ss

By an obvious induction on rules, define which occurrences of variables are
free (or undeclared) and which are bound.  Suppose that $D$ is a variable
declaration, $R$ is a rule, and $S$ is a state of an auxiliary vocabulary.
$R$ is {\em $(D,S)$-perspicuous\/} if it satisfies the following
conditions:
\begin{itemize}

\item no variable is declared (explicitly or implicitly) more than
once in $R$, and

\item $\Bound(R)$ is disjoint from $\Free(R)\cup\Var(D)\cup\Fun(S)$.
\end{itemize}

\paragraph{Programs}
A {\em program\/} is a rule without any undeclared variables.

\subsubsection{Semantics of Rules} 

By induction on $R$, we define the update set $\be=\Updates(D,R,S)$
generated by a rule $R$ at an appropriate state $S$ under a declaration $D$
that covers $R$.  To fire $R$ at $S$ under $D$, fire $\be$.

We stipulate that an arbitrary rule $R$ is equivalent, for given $D$ and
$S$, to a $(D,S)$-perspicuous rule $R'$ obtained from $R$ by renaming the
bound variables.  The equivalence means that $\Updates(D,R,S)
=\Updates(D,R',S)$.

It remains to define $\be=\Updates(D,R,S)$ in the case when $R$ is
$(D,S)$-perspicuous.

If $D$ is not empty, then $\be$ is the union of $\Updates(\emptyset,R,S')$,
where $S'$ ranges over expansions of $S$ such that $\Fun(S')
=\Fun(S)\cup\Var(D)$ and $S'$ is consistent with $D$ (so that the values of
$D$ variables are within their ranges in $S'$).  Notice that $\be
=\emptyset$ if the range of any $D$ variable is empty.

Suppose $D=\emptyset$.  If $R$ is an update instruction then $\be
=\Updates(R,S)$.  If $R$ is a sequence of rules $R_1,\ldots,R_k$, then
$\be$ is the union of the update sets $\Updates(\emptyset,R_i,S$).  Suppose
that $R$ is the conditional rule with clauses $(g_0,R_0),\ldots,(g_k,R_k)$.
Since $R$ is covered by the empty declaration, the guards $g_i$ have no
free variables.  We have two cases as usual.  If all guards $g_i$ fail in
$S$, then $\be$ is empty, and if $g_i$ is the first guard that holds in $S$
then $\be =\NUpdates(\emptyset,R_i,S)$.  Finally, if $R$ is a declaration
rule with declaration $d$ and body $R'$ then $\be =\Updates(d,R',S)$.

\paragraph{Remark}
Suppose that $D=\emptyset$ and $R$ is a sequence of a declaration-free rule
$R_1$ and a declaration rule $R_2$ with atomic declaration ``Var $v$ ranges
over $U$'' followed by a declaration-free body $R_2'$.  Further suppose
that $U$ is empty in a state $S$ appropriate for $R$ and thus
$\Updates(D,R_2,S) =\emptyset$.  Then $\Updates(D,R,S)$ equals
$\Updates(D,R_1,S)$ which may be not empty.  Contrary to the situation in
4.1.2, the empty range does not give inconsistency here.  One cannot choose
an element from the empty set, but one can execute a $R_2'(v)$ for every
$v$ in the empty set: just do not execute anything.

\subsection{Importing Elements}

The recursive definition of rules in 5.3 can be extended by import commands
and/or (qualified) choice commands.  The adjustment of the semantic
definition is straightforward.  For the sake of definiteness, consider the
extension by means of the import constructor.  The most important novelty,
in comparison to 3.2.4, is that reserve elements have to be chosen for all
combinations of the values of explicitly declared variables $u$ such that
the scope of the declaration of $u$ properly includes the given import or
choose subrule.  For example, the rule

\ni\parbox{4in}{
\begin{eatab}
Var $u$ ranges over $U$\\
import $v$\\
\> Parent($v$):=$u$\\
endimport
\end{eatab}}

\ni creates a new child for every element of $U$, and of course all these
new children are different.

To reflect the novelty we redefine the domain of the global choice
function.  Suppose that $D$ is a variable declaration, $R$ is a rule
covered by $D$, $S$ is a state of an auxiliary vocabulary appropriate for
$R$, and $R$ is $(D,S)$-perspicuous.  For every bound variable $v$ of $R$,
list all explicitly declared variables $u$ such that either $u$ occurs in
$D$ or $u$ occurs in $R$ and the scope of the declaration of $u$ properly
includes the scope of the declaration of $v$: $u_1,\ldots,u_l$.  (The
adverb properly is there to exclude $v$ from the list.)  Let
$U_1,\ldots,U_l$ be the ranges of $u_1,\ldots,u_l$ in $S$ respectively, and
$\bar{U}_v$ be the Cartesian product $U_1\times\cdots\times U_l$.  The
desired global function $\xi$ assigns different reserve elements to every
pair $(v,\bar{a})$ where $v\in\Bound(R)$ and $\bar{a}\in\bar{U}_v$.

Here is a variant of the example from 3.2.5:

\ni\parbox{4in}{
\begin{eatab}
Var $u$ ranges over $U$\\
extend Nodes with $v_1,v_2$\\
\> if Leaf($u$) then \\
\> \> FirstChild($u$) := $v_1$\\
\> \> SecondChild($u$) := $v_2$\\
\> \> NextSib($v_1$) := $v_2$\\
\> endif
endextend
\end{eatab}}

\subparagraph{Remark}
Should one provide means to say explicitly that the main existential
variable of a given choose rule depends only on such and such of the free
variables of the rule?  Maybe.  But the need for such means has not been
demonstrated yet. 

\subsection{Runs}

Runs are defined as above.

\section{Distributed Evolving Algebras}

In this section we consider multi-agent computations.  We do not suppose
that agents are deterministic or do only a bounded amount of work at each
step.  The program of an agent may be any program described above.

Agents may share functions, and it is convenient \cite{GR} to assume that all
states of all agents share the same carrier; see the end of 3.2.4 in this
connection.

\subsection{The Self Function}

There is an interesting problem of self identification.  It can be
illustrated on the example of the following simple version of Dijkstra's
dining philosophers protocol (which may deadlock).  There are $n$
philosophers, marked with numbers modulo $n$, each equipped with a fork.  A
philosopher $i$ may think (which requires no forks) or eat using his/her
fork and the fork of philosopher $i+1$.  A fork cannot be used by two
philosophers at the same time.

Using functions
\[\Fork_i = \cases{
\mbox{up}   & if the fork of philosopher $i$ is used,      \cr
\mbox{down} & otherwise,                                   \cr}\]
we can write a separate program for each philosopher $i$.  Intuitively,
however, all philosophers use the same program in the protocol.

To solve such problems, we suppose that each agent $a$ is represented by an
element of the common carrier.  For simplicity, we will not distinguish
between an agent and the element that represents the agent.  Further, we
use a special nullary function Self, interpreted differently by different
agents.  An agent $a$ interprets Self as $a$.  Thus function Self allows an
agent to identify itself among other agents.  Self is a logic name and
cannot be the subject of an update instruction.  To make rules sound a
little better for humans, we use some capitalized pronouns, {\em e.g.\/}
Me, as aliases for Self.  Viewing agents as elements of the carrier is
useful for other purposes as well.  For example, it allows us to model the
creation of new agents.

We return to the dining philosophers protocol.  Here is a possible program
(courtesy of Jim Huggins):

\ni\parbox{4in}{
\begin{eatab}
if Mode(Me)=think and Fork(Me)=Fork(Me+1)=down then\\
\> Fork(Me):=up, Fork(Me+1):=up, Mode(Me):=eat\\
elseif Mode(Me)=eat then\\
\> Fork(Me):=down, Fork(Me+1):=down, Mode(Me):=think\\
endif
\end{eatab}}

It may be convenient to suppress the argument Self.  For example, terms
Mode(Me), Fork(Me) and Fork(Me+1) may be treated as nullary functions and
abbreviated, {\em e.g.\/}, as mode, lfork and rfork, so that the rfork function of
philosopher $i$ is the lfork function of philosopher $i+1$ and mode is a
private function.

\subsection{Basic Definition of Distributed Ealgebras}

A {\em distributed ealgebra\/} $\cA$ consists of the following:
\begin{itemize}

\item
A finite indexed set of single-agent programs $\pi_\nu$, called {\em
modules\/}.  The {\em module names\/} $\nu$ are static nullary function
names. 

\item
A vocabulary $\Ups=\Fun(\cA)$ which includes each\ \ $\Fun(\pi_\nu)
-\{\Self\}$\ \ but does not contain Self.  In addition, $\Ups$ contains a
unary function name Mod.

\item
A collection of $\Ups$-states, called {\em initial states\/} of $\cA$,
satisfying the following conditions:

\begin{itemize}\item 
Different module names are interpreted as different elements.  
\item 
There are only finitely many elements $a$ such that, for some module name
$\nu$, $\Mod(a)=\nu$.
\end{itemize}
\end{itemize}

A state $S$ of vocabulary $\Fun(\cA)$ is a state of $\cA$ if it satisfies
the two conditions imposed on initial states.  In applications it may make
sense to restrict further the notion of state of the ealgebra in question.

An element $a$ is an {\em agent\/} at $S$ if there is a module name $\nu$
such that $S\models\Mod(a)=\nu$; the corresponding $\pi_\nu$ is the program
$\Prog(a)$ of $a$, and $\Fun(\pi_\nu)$ is the vocabulary $\Fun(a)$ of $a$.
Agent $a$ is deterministic if $\Prog(a)$ is so.

$\View_a(S)$ is the reduct of $S$ to vocabulary $\Fun(a)-\{\Self\}$
expanded with Self, which is interpreted as $a$.  Think about $\View_a(S)$
as the local state of agent $a$ corresponding to the global state $S$.  (It
is not necessary to define local states via global states; see \cite{GM} for
example.)

An agent $a$ can {\em make a move\/} at $S$ by firing $\Prog(a)$ at
$\View_a(S)$ and changing $S$ accordingly.  As a part of the move, $a$ may
create new agents, {\em e.g.\/}, by importing reserve elements.

To perform a move of a deterministic agent $a$, fire
\[ \Updates(a,S) = \Updates \Big( \Prog(a), \View_a(S) \Big). \]

Runs of a distributed ealgebra are defined below. 

\paragraph{Cooperative Actions}
Consider a simple scenario with agents Sender and Receiver.  If both are in
mode Ready then Sender passes a value $t_1$ to Receiver who stores it at
location $f(t_2)$.  The transaction is atomic (that is, indivisible), but
the Sender does not have access to $f(t_2)$ and the Receiver does not have
access to $t_1$, and thus neither agent is able to perform the transaction.
A special auxiliary agent is needed to do the job, and it may be convenient
to view the auxiliary agent as a team with members Sender and Receiver.
Using functions Member$_1$ and Member$_2$ to specify the members of the
team, we may write the following rule for the team, where Us is an alias
for Self:

\ni\parbox{4in}{
\begin{eatab}
if Mode(Member$_1$(Us))=Mode(Member$_2$(Us))=Ready then\\
\> $f(t_2) := t_1$\\
endif
\end{eatab}}

\ni In a similar way, one may have larger teams.  Depending on need, teams
may or may not be ordered.

\subsection{Generalizations}

\subsubsection{Active Agents}

Alter definition~6.2 as follows: Require that $\Fun(\cA)$ contains an
additional unary relation name Active and that only agents satisfying the
relation Active ({\em active agents\/}) can make moves.  This is
essentially a generalization; the original definition can be seen as a
special case where all agents are active.

The new definition may be convenient, for example, when the initial state
specifies all agents and their programs, and these agents are activated and
deactivated during the evolution.

The same convenience can be achieved without altering the original
definition.  (This may be useful, for example, if you want to prove
something about all distributed ealgebras and wish to restrict attention to
the basic definition without losing generality.)  Here is one way to do
that.  In order to indicate the program of a potential agent without making
it an actual agent, use an auxiliary unary function name $\Mod'$.
$\Active(t)$ can be viewed as an abbreviation for $\Mod(t)=\Mod'(t)$ except
if Active is the subject of an update instruction.

\begin{eatab}
Active(t):=t$_0$
\end{eatab}

\ni can be viewed as an abbreviation for

\ni\parbox{4in}{
\begin{eatab}
if t$_0$\ \ then Mod(t):=Mod$'$(t)\\ 
else Mod(t):=\undef\\ 
endif
\end{eatab}}

\subsubsection{Active Teams}

The generalized definition of distributed ealgebras described in 6.3.1 is
used in this sub-subsection.  The following problem was raised by Dean
Rosenzweig \cite{R}.

Consider a scenario with (agents called) players and (additional agents
viewed as) teams.  Players form a static universe, teams form another
static universe, and each agent is assigned a program once and for all.
Players are activated and deactivated during the evolution.  A team is
supposed to be active if and only if its members are active.  It follows
that activating one player may necessitate the tedious work of activating
many teams.  Is there a simple and elegant way to ensure that every team is
active when and only when all its members are active?

One obvious solution is to make teams active all the time and augment the
program of each team with a guard stating that all the members are active.
A more radical solution is to make the notion of team a part of the logic
of distributed ealgebras.  It will be ensured automatically that a team is
active if and only if all its members are active.  (It may be also required
that the moves made by a player or any team involving the player are
linearly ordered; see the second property of runs in 6.5.1 in this
connection.)  If substantial programming convenience is demonstrated, use
that solution.

The possibilities to pay a lesser price in logic complication for the
advantages of the radical solution will be discussed elsewhere.  In this
connection, Rosenzweig suggested generalizing further the definition of
6.3.1 by letting a possibly compound Boolean term play the role of Active.
For example, $\Active(v)$ may say that either $v$ is a player satisfying an
auxiliary relation Ac or $v$ is a team with all members satisfying Ac.

\subsection{Sequential Runs}

We return to the basic definition of distributed ealgebras in 6.2.

A {\em pure sequential run\/} $\rho$ of an ealgebra $\cA$ is a sequence
$\langle S_n\st n<\kappa\rangle$ of states of $\cA$, where $S_0$ is an
initial state and every $S_{n+1}$ is obtained from $S_n$ by executing a
move of an agent.  The generalization to the case of external functions or
external locations is relatively straightforward.

\paragraph{Stages}
Since $S_i$ may be equal to $S_j$ for some $i\neq j$, it may be convenient
to speak about stages.  Starting from stage $0$, the run goes through
stages $1$, $2$, {\em etc.\/}.  Formally, stage $i$ can be defined as the pair
$(i,S_i)$.

\paragraph{Quasi-sequential Runs}
An obvious generalization of a sequential run is a quasi-sequential run
$\langle S_n\st n<\kappa\rangle$, where each $S_{n+1}$ is obtained from
$S_n$ by firing a collection $A_n$ of agents.  We do not mean that $A_n$ is
a team; since teams are agents, the definition of sequential runs does not
exclude team moves.  We mean that each $a\in A_n$ makes a move at $S_n$.
If all agents are deterministic, then $S_{n+1}$ is the result of firing
$\ds\bigcup\{\Updates(S,a)\st a\in A_n\}$.

Quasi-sequential runs may arise, for example, if you order moves in real
(physical) time.

\subsection{Partially Ordered Runs}

Partially ordered computations are well known in the literature \cite{L},
\cite{Mz}, \cite{KP}, {\em etc.\/} but we need to define our own version
of that notion for our purposes here.  We restrict attention to the case
where moves are atomic and we use global states.  Non-atomic moves have
been explored in \cite{BGR}.  A simple notion of runs in \cite{GM} does
not use global states.

Let us recall some well known notions.  A {\em poset\/} is a partially
ordered set.  An {\em initial segment\/} of a poset $P$ is a substructure
$X$ of $P$ such that if $x\in X$ and $y<x$ in $P$ then $y\in X$.  Since $X$
is a substructure, $y<x$ in $X$ if and only if $y<x$ in $P$ whenever
$x,y\in X$.  A {\em linearization\/} of a poset $P$ is a linearly ordered
set $P'$ with the same elements such that if $x<y$ in $P$ then $x<y$ in
$P'$.

\subsubsection{Runs}

For simplicity, we restrict attention to pure runs and deterministic
agents.  A {\em run\/} $\rho$ of a distributed ealgebra $\cA$ can be
defined as a triple $(M,A,\sigma)$ satisfying the following conditions
1--4.

\begin{description}
\item[1]
$M$ is a partially ordered set, where all sets $\{y\st y\leq x\}$ are
finite.
\end{description}

Elements of $M$ represent {\em moves\/} made by various agents during the
run.  If $y<x$ then $x$ starts when $y$ is already finished; that explains
why the set $\{y\st y\leq x\}$ is finite.

\begin{description}
\item[2] $A$ is a function on $M$ such that every nonempty set $\{x\st
A(x)=a\}$ is linearly ordered.
\end{description}

$A(x)$ is the agent performing move $x$.  The moves of any single agent are
supposed to be linearly ordered.

\begin{description}
\item[3] $\sigma$ assigns a state of $\cA$ to the empty set and each
finite initial segment of $M$; $\sigma(\emptyset)$ is an initial
state.
\end{description}

$\sigma(X)$ is the result of performing all moves in $X$.

\begin{description}
\item[4] The coherence condition: If $x$ is a maximal element in a finite
initial segment $X$ of $M$ and $Y=X-\{x\}$, then $A(x)$ is an agent in
$\sigma(Y)$ and $\sigma(X)$ is obtained from $\sigma(Y)$ by firing $A(x)$
at $\sigma(Y)$.
\end{description}

Intuitively, a run can be seen as the common part of histories of the same
computation recorded by various observers.  We hope to address this issue
elsewhere.

If agents are not necessarily deterministic, we have to define moves as
state transformers and make the coherence condition more precise:

\begin{description}
\item[4$^*$] If $x$ is a maximal element in a finite initial segment $X$ of
$M$ and $Y=X-\{x\}$, then $A(x)$ is an agent in $\sigma(Y)$, $x$ is a move
of $A(x)$ and $\sigma(X)$ is obtained from $\sigma(Y)$ by performing $x$ at
$\sigma(Y)$.
\end{description}

A run $\rho'$ is an {\em initial segment\/} of a run $\rho$ if (i)~the move
poset of $\rho'$ is an initial segment of the move poset of $\rho$ and
(ii)~the agent and state functions of $\rho'$ are restrictions of those in
$\rho$.  A run $\rho'$ is a {\em linearization\/} of $\rho$ if the move
poset of $\rho'$ is a linearization of that of $\rho$, the agent function
of $\rho'$ is that of $\rho$, and the state function of $\rho'$ is a
restriction of that of $\rho$.  Linearizations are sequential runs.  A
state $S$ of is {\em reachable\/} in a run $\rho$ if it belongs to the
range of the state function of $\rho$.

\begin{Corollary}
All linearizations of the same finite initial segment of $\rho$ have the same
final state.
\end{Corollary}

\begin{Corollary}
A property holds in every reachable state of a run $\rho$ if and only if it
holds in every reachable state of every linearization of $\rho$.
\end{Corollary}

\subsection{Real-time Computations}

Real-time semantics appears in \cite{BGR}.  Ealgebras with clocks made their
debut in \cite{GM}.  We will have to address the issue of real time elsewhere.

\end{document}